\documentclass[letterpaper,twocolumn,conference]{IEEEtran}
\usepackage{helvet}
\usepackage{courier}
\usepackage[utf8]{luainputenc}
\usepackage{float}
\usepackage{amsmath}
\usepackage{amsthm}
\usepackage{amssymb}
\usepackage{graphicx}

\makeatletter

\pdfpageheight\paperheight
\pdfpagewidth\paperwidth

\providecommand{\tabularnewline}{\\}
\floatstyle{ruled}
\newfloat{algorithm}{tbp}{loa}
\providecommand{\algorithmname}{Algorithm}
\floatname{algorithm}{\protect\algorithmname}

\theoremstyle{plain}
\newtheorem{thm}{\protect\theoremname}
\theoremstyle{definition}
\newtheorem{defn}[thm]{\protect\definitionname}
\theoremstyle{plain}
\newtheorem{prop}[thm]{\protect\propositionname}

\usepackage{times}
\usepackage{algorithm}
\usepackage{algpseudocode}
\usepackage{url}
\usepackage{times}
\usepackage{bm}
\usepackage{amsfonts}
\usepackage{booktabs}
\usepackage{subfigure}
\usepackage{cases}
\usepackage{graphicx}
\DeclareMathOperator*{\argmax}{argmax}

\makeatother

\providecommand{\definitionname}{Definition}
\providecommand{\propositionname}{Proposition}
\providecommand{\theoremname}{Theorem}

\begin{document}

\title{Applying DCOP to User Association Problem in Heterogeneous Networks
with Markov Chain Based Algorithm}

\author{\IEEEauthorblockN{Peibo Duan\IEEEauthorrefmark{1}, Guoqiang Mao\IEEEauthorrefmark{1}\IEEEauthorrefmark{2}\IEEEauthorrefmark{3}\IEEEauthorrefmark{4},
Changsheng Zhang\IEEEauthorrefmark{5} and Bin Zhang\IEEEauthorrefmark{5}}\IEEEauthorblockA{\IEEEauthorrefmark{1}School of Computing and Communications\\
 University of Technology, Sydney, Australia}\IEEEauthorblockA{\IEEEauthorrefmark{2}Data61 Australia}\IEEEauthorblockA{\IEEEauthorrefmark{4}Beijing Unriversity of Posts and Telecommunications,
Beijing, China}\IEEEauthorblockA{\IEEEauthorrefmark{4}School of Electronic Information Communications,
Huazhong Unriversity of Science Technology, Wuhan, China}\IEEEauthorblockA{\IEEEauthorrefmark{5}School of Computer Application\\
 Northeastern University\\
 Bin Zhang is the correspondence author }}
\maketitle
\begin{abstract}
Multi-agent systems (MAS) is able to characterize the behavior of
individual agent and the interaction between agents. Thus, it motivates
us to leverage the distributed constraint optimization problem (DCOP),
a framework of modeling MAS, to solve the user association problem
in heterogeneous networks (HetNets). Two issues we have to consider
when we take DCOP into the application of HetNet including: (i) How
to set up an effective model by DCOP taking account of the negtive
impact of the increment of users on the modeling process (ii) Which
kind of algorithms is more suitable to balance the time consumption
and the quality of soltuion. Aiming to overcome these issues, we firstly
come up with an ECAV-$\eta$ (Each Connection As Variable) model in
which a parameter $\eta$ with an adequate assignment ($\eta=3$ in
this paper) is able to control the scale of the model. After that,
a Markov chain (MC) based algorithm is proposed on the basis of log-sum-exp
function. Experimental results show that the solution obtained by
DCOP framework is better than the one obtained by the Max-SINR algorithm.
Comparing with the Lagrange dual decomposition based method (LDD),
the solution performance has been improved since there is no need
to transform original problem into a satisfied one. In addition, it
is also apparent that the DCOP based method has better robustness
than LDD when the number of users increases but the available resource
at base stations are limited. 
\end{abstract}

\section{Introduction}

of agent and the cooperation between agents in Multi-agent System
(MAS), a framework, named distributed constraint optimization problem
(DCOP) in terms of constraints that are known and enforced by distinct
agents comes into being with it. In last decade, the research effort
of DCOP has been dedicated on the following three directions: 1) the
development of DCOP algorithms which are able to better balance the
computational complexity and the accuracy of solution, such as large
neighborhood search method \cite{2015_large_neighborhood_search_quality_guarantee}; 
Markov Chain Monte Carlo sampling method \cite{2016_MCMC_GPUs} \cite{2015_energy}
and distributed junction tree based method \cite{2014_djao} 2) the
extension of classical DCOP model in order to make it more flexible
and effective for practical application, such as expected regret DCOP
model \cite{2016_ER_DCOP}, multi-variable agent decomposition model
\cite{2016_multi_variable_decomposition} and dynamic DCOP model \cite{2011_dynamic_dcop}
3) the application of DCOP in modeling environmental systems, such
as sensor networks \cite{2012_sensor_network,2009_graph,2006_wsn06,2008_robust},
disaster evacuation \cite{2013_disaster_evacuation}, traffic control
\cite{2014_traffic_control,2013_road} and resource allocation \cite{2015_water_resources,2016_Energy_efficient_Smart_Environment,2015_energy}.
In this paper, we take more attention to the application of DCOP.
More precisely, we leverage DCOP to solve user association problem
in the downlink of multi-tier heterogeneous networks with the aim
to assign mobile users to different base stations in different tiers
while satisfying the QoS constraint on the rate required by each user.

is generally regarded as a resource allocation problem \cite{2011_RB,2014_resource_allocation,2016_survey_hetnet,2015_LDD}
in which the resource is defined by the resource blocks (RBs). In
this case, the more RBs allocated to a user, the larger rate achieved
by the user. The methods to solve the user association problem are
divided into centralized controlled and distributed controlled. With
regard to the centralized way, a central entity is set up to collect
information, and then used to decide which particular BS is to serve
which user according to the collected information. A classical representation
of centralized method is Max-SINR \cite{2012_max_signal}. Distributed
controlled methods attract considerable attention in last decade since
they do not require a central entity and allow BSs and users to make
autonomous user association decisions by themselves through the interaction
between BSs and users. Among all available methods, the methods based
on Lagrange dual decomposation (LDD) \cite{2015_LDD} and game theory
\cite{2014_game_theory} have better performance. Hamidreza and Vijay
\cite{2015_LDD} put forward a unified distributed algorithm for cell
association followed by RBs distribution in a $k$-tier heterogeneous
network. With aid of LDD algorithm, the users and BSs make their respective
decisions based on local information and a global QoS, expressed in
terms of minimum achievable long-term rate, is achieved. However,
the constraint relaxation and the backtrack in almost each iteration
are needed to avoid overload at the BSs. In addition, as we will show
later, the number of out-of-service users will increase since a user
always selects a best-rate thereby taking up a large number of RBs
and leaving less for others. Nguyen and Bao \cite{2014_game_theory}
proposed a game theory based method in which the users are modeled
as players who participate in the game of acquiring resources. The
best solution is the one which can satisfy Nash equilibrium (NE).
Loosely speaking, such solution is only a local optima. In addition,
it is difficult to guarantee the quality of the solution.

there is no research of modeling user association problem as MAS.
However, some similar works have been done focusing on solving the
resource management problem in the field of wireless networks or cognitive
radio network \cite{2012_channel_allocation,2012_mas_channel_allocation,2007_resouce_management_cognitive_network}
by DCOP framework. These methods can not be directly applied to user
association problem mainly due to the scale of the models for these
practical applications is relatively small. For instance, Monteiro
\cite{2012_mas_channel_allocation} formalized the channel allocation
in a wireless network as a DCOP with no more than 10 agents considered
in the simulation parts. However, the amount of users and resource
included in a HetNet is always hundreds and thousands. In this case,
a good modeling process along with a suitable DCOP algorithm is necessary.
According to the in-depth analysis above, it motivates us to explore
a good way to solve user association problem by DCOP. The main contributions
of this paper are as follows:
\begin{itemize}
\item An ECAV (Each Connection As Variable) model is proposed for modeling
user association problem using DCOP framework. In addition, we introduce
a parameter $\eta$ with which we can control the scale (the number
of variables and constriants) of the ECAV model. 
\item A DCOP algorithm based on Markov chain (MC) is proposed which is able
to balance the time consumption and the quality of the solution. 
\item The experiments are conducted which show that the results obtained
by the proposed algorithm have superior accuracy compared with the
Max-SINR algorithm. Moreover, it has better robustness than the LDD
based algorithm when the number of users increases but the available
resource at base stations are limited. 
\end{itemize}
The rest of this paper is organized as follows. In Section \ref{preliminary},
the definition of DCOP and the system model of user association problem
along with its mixed integer programming formulation are briefly introduced.
In Section \ref{formulation_with_DCOP}, we illustrate the ECAV-$\eta$
model. After that, a MC based algorithm is designed in section \ref{markov_chain_algorithm}.
We explore the performance of the DCOP framework by comparing with
the Max-SINR and LDD methods in Section \ref{experimental_evaluation}.
Finally, Section \ref{conclusion} draws the conclusion.

\section{Preliminary}

\label{preliminary}

This section expounds the DCOP framework and system model of user
association problem along with its mixed integer programming formulation.

\subsection{DCOP}

\label{dcop}

The definitions of DCOP have a little difference in different literatures
\cite{2011_three_tuple_DCOP,2014_djao,2016_ER_DCOP} \footnote{\cite{2011_three_tuple_DCOP} formalized the DCOP as a three tuples
model. \cite{2014_djao} adopted a four-tuples model while \cite{2016_ER_DCOP}
used a five tuples model}. In this paper, we formalize the DCOP as a four tuples model $<\mathcal{A},\mathcal{V},\mathcal{D},\mathcal{C}>$
where $\mathcal{A}=\{a_{1},a_{2},...,a_{\lvert\mathcal{A}\rvert}\}$
consists of a set of agents, $\mathcal{V}=\{v_{1},v_{2},...,v_{n}\}$
is the set of variables in which each variable $v_{i}\in\mathcal{V}$
only belongs to an agent $a\in\mathcal{A}$. Each variable has a finite
and discrete domain where each value represents a possible state of
the variable. All the domains of different variables consist of a
domain set $\mathcal{D}=\{d_{1},d_{2},...,d_{n}\}$, where $d_{i}$
is the domain of $v_{i}$. A constraint $c\in\mathcal{C}=\{c_{1},c_{2},...,c_{\lvert\mathcal{C}\rvert}\}$
is defined as a mapping from the assignments of $m$ variables to
a positive real value:

\begin{equation}
R(c):d_{i_{1}}\times d_{i_{2}}\times\cdots d_{i_{m}}\rightarrow\mathbb{R}^{+}\label{dcop_definition_constraint}
\end{equation}

The purpose of a DCOP is to find a set of assignments of all the variables,
denoted as $\mathcal{X}^{*}$, which maximize the utility, namely
the sum of all constraint rewards:

\begin{equation}
\argmax_{\mathcal{X}^{*}}{\sum_{\mathcal{C}}R(c)}\label{formula_optimization_function}
\end{equation}

\subsection{System Model of User Association Problem}

\label{system_model}

\begin{figure*}
	\centering
	\subfigure[A simple instance of HetNet]{
		\label{ecav_instance:a} 
		\includegraphics[width=3in]{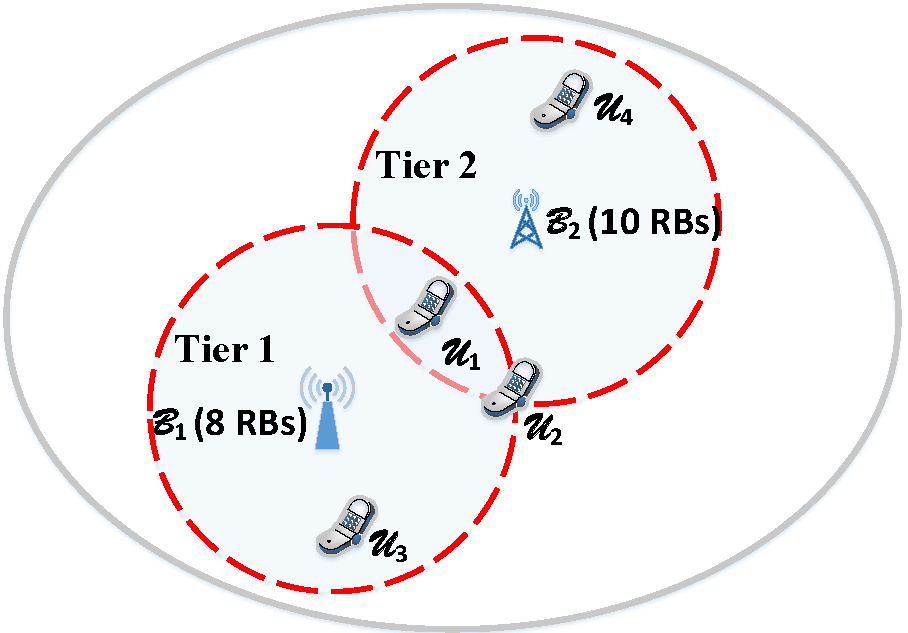}}
	\hspace{0.5in}
	\subfigure[ECAV model]{
		\label{ecav_instance:b} 
		\includegraphics[width=2.2in]{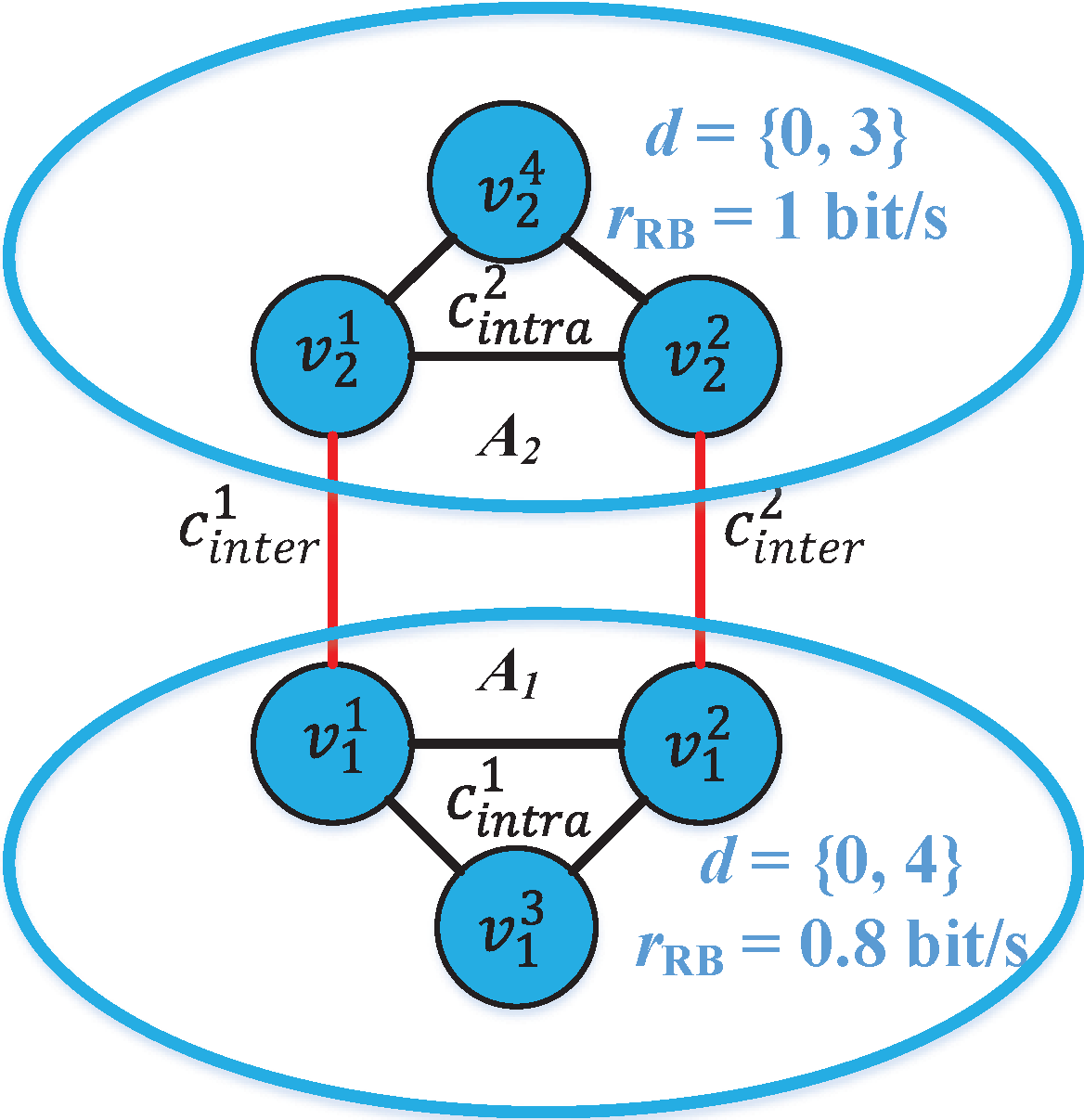}}
	\caption{A simple instance of HetNet with ECAV modeling}
	\label{ecav_instance} 
\end{figure*}

Consider a $k$-tier HetNet where all the BSs in the same tier have
the same configurations. For example, a two-tier network including
a macro BS ($\mathcal{B}_{1}$) and a femto BS, ($\mathcal{B}_{2}$),
is shown in Fig.\ref{ecav_instance}. The set of all BSs is denoted
as $\mathcal{B}=\{\mathcal{B}_{1},\mathcal{B}_{2},...,\mathcal{B}_{\mathcal{NB}}\}$
where $\mathcal{NB}$ is the total number of BSs. All the BSs in the
$k$th tier transmit with the same power $P_{k}$. The total number
of users is denoted by $\mathcal{NU}$ and the set of all users is
$\mathcal{U}=\{\mathcal{U}_{1},\mathcal{U}_{2},...,\mathcal{U}_{\mathcal{NU}}\}$.

With OFDMA technology in LTE-Advanced networks, the resource, time-frequency,
is divided into blocks where each block is defined as a resource block
(RB) including a certain time duration and certain bandwidth \cite{2011_RB}.
In this paper, the resource configured at each BS is in the format
of RB so that its available RBs are decided by the bandwidth and the
scheduling interval duration allocated to that BS. We assume the BSs
in the HetNet share the total bandwidth such that both intra- and
inter-tier interference exist when the BSs allocate RBs to the users
instantaneously.

Assuming the channel state information is available at the BSs, the
${\rm SINR}$ experienced by user $\mathcal{U}_{j}$, served by $\mathcal{B}_{i}$
in the $k$th tier is given by

\begin{equation}
{\rm SINR}_{ij}=\frac{P_{k}g_{ij}}{\sum_{\mathcal{B}_{l}\in\mathcal{B}/\{\mathcal{B}_{i}\}}{P_{k}g_{ij}}+BN_{0}}\label{sinr}
\end{equation}

In (\ref{sinr}), $g_{ij}$ is the channel power gain between $\mathcal{U}_{j}$
and $\mathcal{B}_{i}$, $\mathcal{B}/\{\mathcal{B}_{i}\}$ represents
all the BSs in $\mathcal{B}$ except $\mathcal{B}_{i}$, $B$ is the
bandwidth and $N_{0}$ is noise power spectral density. The channel
power gain includes the effect of both path loss and fading. Path
loss is assumed to be static and its effect is captured in the average
value of the channel power gain, while the fading is assumed to follow
the exponential distribution.

From the above, the efficiency of user $\mathcal{U}_{j}$ powered
by BS $\mathcal{B}_{i}$, denoted as $e_{ij}$, is calculated as

\begin{equation}
e_{ij}=log_{2}(1+{\rm SINR}_{ij})\label{unit_efficiency}
\end{equation}

Given the bandwidth $B$, time duration $T$ and the scheduling interval
$\varGamma$ configured at each RB, we attain the unit rate at $\mathcal{U}_{j}$
upon one RB as follows

\begin{equation}
u_{ij}=\frac{BTe_{ij}}{\varGamma}\label{rate_RB}
\end{equation}

On the basis of formula (\ref{rate_RB}), the rate received at $\mathcal{U}_{j}$
with $n_{ij}$ RBs provided by $\mathcal{B}_{i}$ in the $k$th tier
is

\begin{equation}
r_{ij}=n_{ij}u_{ij}\label{rate}
\end{equation}

Associated with each user is a quality-of-service (QoS) constraint.
This is expressed as the minimum total rate the user should receive.
Denoting the rate requiremnt of the $j$th user by $\gamma_{j}$,
the minimum number of RBs required to satisfy $\gamma_{j}$ is calculated
by:

\begin{equation}
n_{min}^{ij}=\lceil\frac{\gamma_{j}}{u_{ij}}\rceil\label{unit_RB}
\end{equation}

\noindent in which $\lceil\cdot\rceil$ is a ceiling function.

\subsection{Mixed Integer Programming Formulation}

The formulations of user association problem by mixed linear programming
are similar in a series of papers (see the survey literature \cite{2016_survey_hetnet}).
in this paper, we present a more commonly used formulation as follows

\begin{subequations} 
\begin{align}
{\rm maximize}\quad & \mathcal{F}=\sum_{i\in\mathcal{B}}{\sum_{j\in\mathcal{U}}}x_{ij}r_{ij}\label{formula_objective}\\
\mbox{s.t.} & \sum_{i\in\mathcal{B}}{x_{ij}r_{ij}}\geqslant\gamma,\forall\mathcal{U}_{j}\in\mathcal{U}\label{formula_rate_constraint}\\
 & \sum_{j\in\mathcal{U}}{x_{ij}n_{ij}}\leqslant N_{i},\forall\mathcal{B}_{i}\in\mathcal{B}\label{formula_rb_constraint}\\
 & \sum_{i\in\mathcal{B}}{x_{ij}}\leqslant1,\forall\mathcal{U}_{j}\in\mathcal{U}\label{formula_link_constraint}\\
 & n_{ij}\in\{0,1,...,N_{i}\},\forall\mathcal{B}_{i}\in\mathcal{B},\forall\mathcal{U}_{j}\in\mathcal{U}\label{formula_nrb_constraint}\\
 & x_{ij}\in\{0,1\},\forall\mathcal{B}_{i}\in\mathcal{B},\forall\mathcal{U}_{j}\in\mathcal{U}\label{formula_vector_constraint}
\end{align}
\end{subequations}

The first constraint ensures the rate QoS requirement from users.
Constraint (\ref{formula_rb_constraint}) indicates that the amount
of RBs consumed at the same BS is no more than the total RBs $N_{i}$
configurated at the BS. Constraint (\ref{formula_link_constraint})
guarantees one user associated with a unique BS. Constraint (\ref{formula_nrb_constraint})
guarantees the number of RBs a BS allocates to a user falls within
the range from zero and $N_{i}$. The last constraint (\ref{formula_vector_constraint})
guarantees the connection between a user and a BS has two states denoted
by a binary variable. The objective function (\ref{formula_objective})
refers to the sum of rate rather than a function acted on the rate
such as $U(x_{ij}r_{ij})$ (e.g. $U(x)=log(1+x)$) in some references.
Generally, two phases are needed to gain the solution including: 1)
transforming original problem into a satisfied one through relaxing
Constraint (\ref{formula_nrb_constraint}) by $n_{ij}\in\{0,n_{min}^{ij}\}$;
2) the left RBs in each BS will be allocated to users in order to
maximize the objective function.

\section{Formulation with DCOP}

\label{formulation_with_DCOP}

In this section, we expound and illustrate the ECAV model along with
its modified version ECAV-$\eta$.

\subsection{ECAV Model}

Before giving the formulation based on DCOP, we firstly introduce
the definition of candidate BS:
\begin{defn}
\label{candidate_BS} we declare $\mathcal{B}_{i},i\in\mathcal{B}$
is a candidate BS of $\mathcal{U}_{j},j\in\mathcal{U}$ if the rate
at $\mathcal{U}_{j}$ is above the threshold $\gamma$ with $n_{min}^{ij}$
RBs provided by $\mathcal{B}_{i}$. Simultaneously, $n_{min}^{ij}$
should be less than the total number of RBs ($N_{i}$) configurated
at $\mathcal{B}_{i}$. 
\end{defn}
After confirming the set of candidate BSs of $\mathcal{U}_{j},j\in\mathcal{U}$,
denoted by, $\mathcal{CB}_{j}$, $\mathcal{U}_{j}$ sends messages
to its candidate BSs so that each $\mathcal{B}_{i},i\in\mathcal{NB}$
gets knowledge of its possible connected users. We define each possible
connection between $\mathcal{U}_{j}$ and its candidate BS $\mathcal{B}_{i}$
as a variable, denoted by $\mathcal{V}_{i}^{j}$. In this case, all
the variables are divided into $\mathcal{NB}$ groups according to
the potential connection between users and different BSs. The domain
of each variable $\mathcal{V}_{i}^{j}$, denoted by $\mathcal{D}_{i}^{j}=\{0,n_{min}^{ij},...,N_{i}\}$,
where $\mathcal{V}_{i}^{j}=0$ if no RB is allocated to $\mathcal{U}_{j}$,
otherwise, $\mathcal{V}_{i}^{j}\geqslant n_{min}^{ij}$. We define
each group as an agent. Thus, an $n$-ary constraint exists among
$n$ variables (intra-constraint) to guarantee that there is no overload
at $\mathcal{B}_{i}$. Note that a user may have more than one candidate
BS, there are constraints (inter-constraints) connecting the variables
affiliated to different agents on account of the assumption that a
unique connection exists between a user and a BS. Generally speaking,
the utility (objective) function in the DCOP model is the sum of constraint
rewards which reflects the degree of constraint violations. We define
the reward $R(c)$ of inter- and intra-constraints in the ECAV model
as follows. For $\forall c\in\mathcal{C}_{inter}$

\begin{subnumcases}{R(c)=} \label{ecav_constraint_1_b} -\infty,
& $\exists\mathcal{V}_{i}^{j_{1}},\mathcal{V}_{i}^{j_{2}}\in\psi(c),Val(\mathcal{V}_{i}^{j_{1}/j_{2}})>0$
\\
 \label{ecav_constraint_1_a} 0, & Otherwise

\end{subnumcases}

\noindent For $\forall c\in\mathcal{C}_{intra}$

\begin{subnumcases}{R(c)=} \label{ecav_constraint_2_a} -\infty,
& $\sum_{\mathcal{V}_{i}^{j}\in\psi_{c}}{Val(\mathcal{V}_{i}^{j})>N_{i}}$
\\
 \label{ecav_constraint_2_b} \sum{\mathcal{V}ji \in \psi(c)}
{r{ij}}, & otherwise \end{subnumcases}

In constraint (\ref{ecav_constraint_1_b}), $\psi(c)$ is the subset
of variables connected by constraint $c$. $Val(\mathcal{V}_{i}^{j})$
represents the assignment of $\mathcal{V}_{i}^{j}$. A reward (we
use $-\infty$ in this paper) is assigned to the constraints if there
at least two variables are non-zero at the same time (unique connection
between a user and a BS). Otherwise, the reward is equal to zero.
In constraint (\ref{ecav_constraint_2_a}), the reward is $-\infty$
once there is a overload at the BS. Otherwise, the reward is the sum
of the rates achieved at users.

It is easy to find that a variable in the ECAV model with non-zero
assignment covers constraint (\ref{formula_rate_constraint}) and
(\ref{formula_nrb_constraint}) in the mixed integer programming formulation.
Moreover, intra and inter-constraints respectively cover constraint
(\ref{formula_rb_constraint}) and constraint(\ref{formula_link_constraint}).
Therefore, The global optimal solution $\mathcal{X}^{*}$ obtained
from the ECAV model is consistent with the one obtained from the mixed
integer programming formulation, denoted as $\mathcal{X}$ \footnote{We say $\mathcal{X}^{*}$ is consistent with $\mathcal{X}$ when the
total rate calculated by objective function (\ref{formula_optimization_function})
and (\ref{formula_objective}) is equal. This is because there may
be no more than one optimal solution.}.

To better understand the modeling process, we recall the instance
in Fig.\ref{ecav_instance} where the candidate BSs of $\mathcal{U}_{1}$
and $\mathcal{U}_{2}$ are the same, denoted as $\{\mathcal{B}_{1},\mathcal{B}_{2}\}$,
while the candidate BSs of $\mathcal{U}_{3}$ and $\mathcal{U}_{4}$
are respectively $\{\mathcal{B}_{1}\}$ and $\{\mathcal{B}_{2}\}$.
We assume the total RBs configurated at $\mathcal{B}_{1}$ and $\mathcal{B}_{2}$
is 8 and 10. For simplicity, we assume the rate of each user served
by one RB provided by $\mathcal{B}_{1}$ is 0.8 bit/s. And 1 bit/s
of each user is served by $\mathcal{B}_{2}$. Then, the ECAV model
is shown in Fig.\ref{ecav_instance:b}. There are two agents named
$\mathcal{A}_{1}$ and $\mathcal{A}_{2}$. The variables in $\mathcal{A}_{1}$
are $\mathcal{V}_{1}^{1},\mathcal{V}_{1}^{2}$ and $\mathcal{V}_{1}^{3}$
where $\mathcal{V}_{i}^{j}$ refers to a connection between user $\mathcal{U}_{j}$
and $\mathcal{B}_{i}$. Similarly, the variables in $\mathcal{A}_{2}$
are $\mathcal{V}_{2}^{1},\mathcal{V}_{2}^{2}$ and $\mathcal{V}_{2}^{4}$.
Assuming the threshold rate is 3 bit/s, we can calculate that at least
$\lceil\frac{3}{0.8}\rceil=4$ RBs needed for the users served by
$\mathcal{B}_{1}$, thus the domain of each variable in $\mathcal{A}_{1}$
is $\{0,4,...,8\}$. Also, the domain of each variable in $\mathcal{A}_{2}$
is $\{0,3,...,10\}$. The black lines in each agent are two 3-nry
intra-constraints, thus $\mathcal{C}_{intra}=\{\mathcal{C}_{intra}^{1},\mathcal{C}_{intra}^{2}\}$.
The red lines connecting two agents are two intra-constraints, thus
$\mathcal{C}_{inter}=\{\mathcal{C}_{inter}^{1},\mathcal{C}_{inter}^{2}\}$.
We use $\mathcal{C}_{intra}^{1}$ and $\mathcal{C}_{inter}^{1}$ to
illustrate how the reward of constraint works in different conditions.
Considering $\mathcal{C}_{intra}^{1}$, the reward is $-\infty$ when
all the variables associated with $\mathcal{C}_{intra}^{1}$ have
the same assignment 4. Thus the total number of RBs consumed by three
users is 12 which is more than 8 RBs configurated at $\mathcal{B}_{1}$.
Otherwise, the reward is 0.8 $\times$ 4 $\times$ 3 = 9.6 (bit/s)
calculated according to (\ref{rate}). Considering $\mathcal{C}_{inter}^{1}$,
the reward is $-\infty$ when the assignment of $\mathcal{V}_{2}^{1}$
is 3 and the assignment of $\mathcal{V}_{1}^{1}$ is 4 because it
means $\mathcal{U}_{2}$ will connect with more than one BSs ($\mathcal{B}_{1}$
and $\mathcal{B}_{2}$), which violates the assumption of unique connection.
Otherwise, the reward is 0 (\ref{ecav_constraint_1_a})). If there
is no constraint violated, the final utility calculated by the objective
function is the total rate in the whole HetNet (constraint (\ref{ecav_constraint_2_b})).

\subsection{ECAV-$\eta$ Model}

The scale of an ECAV model, referring to the number of agents and
constraints, is related to the number of users, BSs and the candidate
BSs hold at each user. However, some candidate BSs of the user can
be ignored because these BSs are able to satisfy the requirement of
the user but with massive RBs consumed. It can be illustrated by the
number of RBs a BS allocate to a user is inversely proportional to
the geographical distance between them. In this way, we introduce
a parameter $\eta$ with which we limit the number of candidate BSs
of each user is no more than $\eta$. The following algorithms present
the selection of top $\eta$ candidate BSs (denoted by $\hat{\mathcal{CB}}$)
and the modeling process of ECAV-$\eta$.

\begin{algorithm}[htbp]
\caption{$\hat{\mathcal{CB}_{j}}$ of user $\mathcal{U}_{j},j\in\mathcal{U}$
based on $\eta$}
\label{alg1} 
\begin{algorithmic}

\State{\textbf{Input:} The information of HetNet ($\mathcal{B}$,
$\mathcal{U}$, $\gamma$, $\eta$)}

\State{\textbf{Output:} The set of candidate BS $\hat{\mathcal{CB}}$
based on $\eta$}

\State{\textbf{Initialize:} $\mathcal{CB}_{j}\leftarrow\phi$, $\hat{\mathcal{CB}_{j}}\leftarrow\phi$}

\Procedure {GetALLCandidateBS}{}

\For{$i\in\mathcal{B},j\in\mathcal{U}$} \label{alg1AllCSStart}

\If{$r_{ij}\geq\gamma$}

\State $\mathcal{CB}_{j}\bigcup\{\mathcal{B}_{i}\}$

\EndIf

\EndFor \label{alg1AllCSEnd}

\EndProcedure

\Procedure{GetPartialCandidateBS}{}

\State BubbleSort ($\mathcal{CB}_{j}$) \label{alg1SortStart} \Comment{sorting
by SINR}

\If{$|\mathcal{CB}_{j}|>\eta$}

\For{$n$ from 1 to $\eta$}

\State $\hat{\mathcal{CB}_{j}^{n}}\leftarrow\mathcal{CB}_{j}^{n}$
\Comment{get $\eta$ candidate BSs}

\EndFor

\Else

\State $\hat{\mathcal{CB}_{j}}\leftarrow\mathcal{CB}_{j}$

\EndIf

\EndProcedure

\Procedure{BubbleSort}{$\mathcal{CB}_{j}$}

\For{$m$ from 1 to $|\mathcal{CB}_{j}|$} \label{alg1SortBySINRStart}

\For{$n$ from $|\mathcal{CB}_{j}|$ to $m+1$}

\If{$SINR(\mathcal{CB}_{j}^{n})>SINR(\mathcal{CB}_{j}^{n-1})$}

\State exchane $\mathcal{CB}_{j}^{n}$ and $\mathcal{CB}_{j}^{n-1}$

\EndIf

\EndFor

\EndFor \label{alg1SortBySINREnd}

\EndProcedure

\end{algorithmic} 
\end{algorithm}

Algorithm \ref{alg1} is the pseudo code for determining $\hat{\mathcal{CB}}$.
It is executed by each user distributely. More precisely, a user estimates
its total candidate BSs $\mathcal{CB}$ by the procedure from line
\ref{alg1AllCSStart} to \ref{alg1AllCSEnd}. Based on \ref{unit_efficiency}
to \ref{unit_RB}, the candidate BSs of a user is ordered according
to the unit number of RBs consumed at such user served by different
BSs (from line \ref{alg1SortBySINRStart} to \ref{alg1SortBySINREnd}).
The time consumption of Algorithm \ref{alg1} mainly consists of two
parts. One is the dermination of $\mathcal{CB}$ with time complexity
$O(\mathcal{NB})$. The other is the ordering operation with time
complexity $O(\mathcal{NB}^{2})$. As a result, the total time expended
of Algorithm \ref{alg1} is $O(\mathcal{NB}+\mathcal{NB}^{2})$. With
$\hat{\mathcal{CB}}$, we present the pseudo code in relation to the
building of ECAV-$\eta$ model.

\begin{algorithm}[htbp]
\caption{ECAV-$\eta$}
\label{alg2} \begin{algorithmic}

\State{\textbf{Initialize:}}

\State $\{\mathcal{A},\mathcal{V},\mathcal{D},\mathcal{R}\}\leftarrow\phi$
\Comment{elements in DCOP model}

\State $R\leftarrow\phi$ \Comment{utility upon each constraint}

\Procedure{SetECAV}{}

\For{$i\in\mathcal{B}$}

\State $\mathcal{A}\bigcup\{\mathcal{A}_{i}\}$ \label{ecav_agent}

\EndFor

\For{$\forall j\in\mathcal{U},m\in\hat{\mathcal{CB}_{j}}$} \label{ecav_user_start}

\State $\mathcal{V}\bigcup\{\mathcal{V}_{i}^{j}\}$

\State $\mathcal{D}_{i}^{j}\leftarrow\{0,n_{min}^{ij}\}$

\State $\mathcal{D}\bigcup\{\mathcal{D}_{i}^{j}\}$

\State $\mathcal{C}\bigcup\{\mathcal{C}_{inter}^{j}\}$

\State $R\bigcup\{R(\mathcal{C}_{inter}^{j})\}$ \Comment{based
on (\ref{ecav_constraint_1_a}), (\ref{ecav_constraint_1_b})}

\EndFor \label{ecav_user_end}

\For{$i\in\mathcal{A}$}

\State $\mathcal{C}\bigcup\{\mathcal{C}_{intra}^{i}\}$ \label{ecav_intra_c_start}

\State $R\bigcup\{R(\mathcal{C}_{intra}^{i})\}$ \Comment{based
on (\ref{ecav_constraint_2_a}), (\ref{ecav_constraint_2_b})} \label{ecav_intra_c_end}

\EndFor

\EndProcedure

\end{algorithmic} 
\end{algorithm}

As for Algorithm \ref{alg2}, it firstly sets up the agents distributely
(line \ref{ecav_agent}). It takes $O(1)$. After that, each user
determines variables, domains as well as inter-constraints from line
\ref{ecav_user_start} to \ref{ecav_user_end}. This is also carried
out in parallel with $O(1)$. Finally, the intra-constraints are constructed
by each agent with $O(1)$ (line \ref{ecav_intra_c_start} to \ref{ecav_intra_c_end}).
The total time complexity is $O(3)$.

\section{Markov Chain based Algorithm}

\label{markov_chain_algorithm}

DCOP, to some degree, is a combinatorial optimization problem in which
the variables select a set of values to maximize the objective function
without or with the minimum constraint violation. We use $S$ to denote
the set of all possible combination of assignments of variables. Also,
we call each element $s\in S$ as a \textbf{candidate solution}. Considering
an ECAV-$\eta$ model in which the four tuples are as follows:
\begin{itemize}
\item $\mathcal{A}=\{\mathcal{A}_{1},\mathcal{A}_{2},...,\mathcal{A}_{\mathcal{NB}}\}$ 
\item $\mathcal{V}=\{\mathcal{V}_{i}^{j}|$ a connection between $\mathcal{U}_{j}$
and $\mathcal{B}_{i}\}$ 
\item $\mathcal{D}=\{\mathcal{D}_{i}^{j}|\mathcal{D}_{i}^{j}=\{0,N_{min}^{ij}\}\}$ 
\item $\mathcal{C}=\mathcal{C}_{inter}\cup\mathcal{C}_{intra}$ 
\end{itemize}
\noindent We are able to rewrite the model in the following way:

\begin{subequations} \label{transformation_1} 
\begin{align}
\max\limits _{s\in S} & \sum_{i\in|\mathcal{B}|,j\in|\mathcal{U}|}Val(\mathcal{V}_{i}^{j})\label{transformation_1_objective_function}\\
s.t. & \nexists\mathcal{V}_{i}^{j_{1}},\mathcal{V}_{i}^{j_{2}}\in\psi(c),Val(\mathcal{V}_{i}^{j_{1}/j_{2}})>0\label{transformation_1_c1}\\
 & \sum_{\mathcal{V}_{i}^{j}\in\psi(c)}{r_{ij}}>N_{min}^{ij}\label{transformation_1_c2}
\end{align}
\end{subequations}

\noindent After that, a convex log-sum-exp approximation of (\ref{transformation_1_objective_function})
can be made by:

\begin{equation}
\max\limits _{s\in S}\sum_{i\in|\mathcal{B}|,j\in|\mathcal{U}|}Val(\mathcal{V}_{i}^{j})\approx\frac{1}{\beta}log(\sum_{s\in S}exp(\beta\sum_{i\in|\mathcal{B}|,j\in|\mathcal{U}|}Val(\mathcal{V}_{i}^{j}))\label{log_sum_exp_approximation}
\end{equation}

\noindent where $\beta$ is a positive constant. We then estimate
the gap between log-sum-exp approximation and (\ref{transformation_1_objective_function})
by the following proposition in \cite{2004_convex_optimization}:
\begin{prop}
\label{gap} Given a positive constant $\beta$ and $n$ nonnegative
values $y_{1},y_{2},...,y_{n}$, we have \\
 
\begin{equation}
\begin{aligned}\max\limits _{i=1,2,...,n}y_{i} & \leq\frac{1}{\beta}log(\sum_{i=1}^{n}exp(\beta y_{i}))\\
 & \leq\max\limits _{i=1,2,...,n}y_{i}+\frac{1}{\beta}logn
\end{aligned}
\label{gap_equation}
\end{equation}

\end{prop}
\noindent In addition, the objective function (\ref{transformation_1_objective_function})
has the same optimal value with the following transformation:

\begin{equation}
\begin{aligned}\max\limits _{p_{s}\geqslant0} & \sum_{s\in S}p_{s}\sum_{i\in|\mathcal{B}|,j\in|\mathcal{U}|}Val_{s}(\mathcal{V}_{i}^{j})\\
s.t. & \sum_{s\in S}p_{s}=1
\end{aligned}
\label{objective_transformation_1}
\end{equation}

\noindent in which $\sum_{i\in|\mathcal{B}|,j\in|\mathcal{U}|}Val_{s}(\mathcal{V}_{i}^{j})$
is the reward with a candidate solution $s$. For simplicity, we use
$g_{\beta}=\frac{1}{\beta}log(\sum_{s\in S}exp(\beta\sum_{i\in|\mathcal{B}|,j\in|\mathcal{U}|}Val(\mathcal{V}_{i}^{j}))$.
Hence, on the basis of formulations (\ref{log_sum_exp_approximation})
and (\ref{gap_equation}), the estimation of (\ref{transformation_1_objective_function})
can be employed by evaluating $g_{\beta}$ in the following way:

\begin{equation}
\begin{aligned}\max\limits _{p_{s}\geqslant0} & \sum_{s\in S}p_{s}\sum_{i\in|\mathcal{B}|,j\in|\mathcal{U}|}Val_{s}(\mathcal{V}_{i}^{j})-\frac{1}{\beta}\sum_{s\in S}p_{s}logp_{s}\\
s.t. & \sum_{s\in S}p_{s}=1
\end{aligned}
\label{objective_transformation_2}
\end{equation}

Assuming $s^{*}$ and $\lambda^{*}$ are the primal and dual optimal
points with zero duality gap. By solving the Karush-Kuhn-Tucker
(KKT) conditions \cite{2004_convex_optimization}, we can obtain the
following equations:

\begin{subequations} \label{KKT} 
\begin{align}
 & \sum_{i\in|\mathcal{B}|,j\in|\mathcal{U}|}Val_{s}(\mathcal{V}_{i}^{j})-\frac{1}{\beta}logp_{s^{*}}-\frac{1}{\beta}+\lambda=0,\forall s\in S\\
 & \quad\ \sum_{s\in S}p_{s^{*}}=1\\
 & \quad\ \lambda\geq0
\end{align}
\end{subequations}

\noindent Then we can get the solution of $p_{s^{*}}$ as follows:

\begin{equation}
p_{s^{*}}=\frac{exp(\beta\sum_{i\in|\mathcal{B}|,j\in|\mathcal{U}|}Val_{s}(\mathcal{V}_{i}^{j}))}{\sum_{s\in S}exp(\beta\sum_{i\in|\mathcal{B}|,j\in|\mathcal{U}|}Val_{s}(\mathcal{V}_{i}^{j}))}
\end{equation}

On the basis of above transformation, the objective is to construct
a MC with the state space being $S$ and the stationary distribution
being the optimal solution $p_{s^{*}}$ when MC converges. In this
way, the assignments of variables will be time-shared according to
$p_{s^{*}}$ and the system will stay in a better or best solution
with most of the time. Another important thing is to design the nonnegative
transition rate $q_{s,s'}$ between two states $s$ and $s'$. According
to \cite{2013_markov_chain_for_network}, a series of methods are
provided which not only guarantee the resulting MC is irreducible,
but also satisfy the balance equation: $p_{s}q_{s,s'}=p_{s'}q_{s',s}$.
In this paper, we use the following method:

\begin{equation}
q_{s,s'}=\alpha[exp(\beta\sum_{i\in|\mathcal{B}|,j\in|\mathcal{U}|}Val_{s}(\mathcal{V}_{i}^{j}))]^{-1}\label{transit_rate}
\end{equation}

The advantage of (\ref{transit_rate}) is that the transition rate
is independent of the performance of $s'$. A distributed algorithm,
named Wait-and-Hp \footnote{to save space, we advise readers to get more details from literature
\cite{2013_markov_chain_for_network}} in \cite{2013_markov_chain_for_network}, is used to get the solution
after we transform DCOP into a MC. However, as the existence of inter-
and intra- constriants in, a checking through the way of message passing
is made in order to avoid constraint violation.

\section{Experimental Evaluation}

\label{experimental_evaluation}

\subsection{Experimental Setting}

In this section, we test the performance of the MC based algorithm
with different assginments of $\eta$ in the ECAV model. A simulated
environment including a three-tiers HetNet created within a $1000m\times1000m$
square is considered. In the system, there is one macro BS, 5 pico
BSs and 10 femto BSs with their transmission powers respectively 46,
35, and 20 dBm. The macro BS is fixed at the center of the square,
and the other BSs are randomly distributed. The path loss between
the macro (pico) BSs and the users is defined as $L(d)=34+40log_{10}(d)$,
while the pass loss between femto BSs and users is $L(d)=37+30log_{10}(d)$.
The parameter $d$ represents the Euclidean distance between the BSs
and the users in meters. The noise power refers to the thermal noise
at room temperature with a bandwidth of 180kHz and equals to -111.45
dBm. One second scheduling interval is considered. Without special
illustration, 200 RBs are configured at macro BS, as well as 100 and
50 RBs are configured at each pico and femto BS. In addition, all
the results are the mean of 10 instances.

\subsection{Experimental Results}

\begin{figure}[htbp]
\centering{} \includegraphics[scale=0.41]{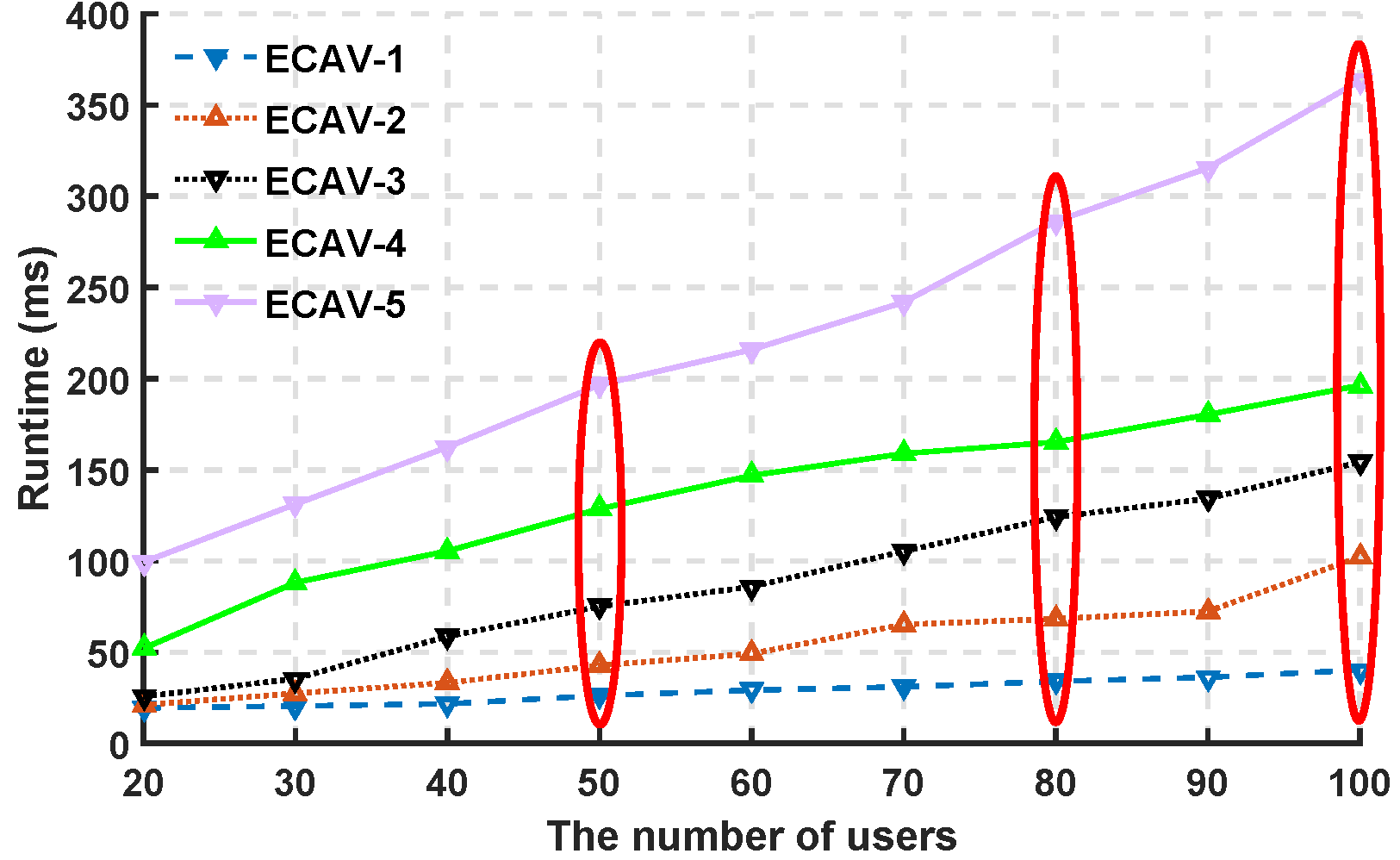} \caption{The runtime of ECAV-$\eta$ ($\eta=1-5$) with different number of
users in the HetNet \label{runtime}}
\end{figure}

We firstly discuss the impact of different assignments of $\eta$
on the performance of ECAV model from the point of view of the runtime
and the quality of solution. More precisely, we generate different
number of users ranging from 20 to 100 with the step interval of 10.
The time consumed by the MC based algorithm is displayed in Fig.\ref{runtime}.
It is clear to see that more time is needed when the number of users
increases. Also, the growth of runtime is depended on the value of
$\eta$. Specially, there is an explosive growth of runtime when we
set $\eta$ from four to five. As previously stated, this is caused
by more candidate BSs considered by each user. However, the quality
of the solutions with different values of $\eta$ is not obviously
improved according the results in Table \ref{table_model_variables}.
For instance, the average rate achieved at each user is only improved
no more than 0.1 bit/s when the number of users are 100 with the values
of $\eta$ are 3 and 5. It is difficult to make a theoretical analysis
of the realationship between $\eta$ and the quality of the solution.
We leave this research in future works.

From above analysis, we set $\eta=3$ in the following experiments
in order to balance the runtime and performance of the solution. In
addition, we test the performance of the MC based algorithm comparing
with its counterparts Max-SINR and LDD based algorithms.

\begin{table}[htbp]
\global\long\def\arraystretch{1.3}
 \caption{The average rate (bit/s) achieved at each user}
\label{table_model_variables} \centering %
\begin{tabular}{cccccc}
\hline 
\textbf{Users}  & \textbf{$\eta=1$}  & \textbf{$\eta=2$}  & \textbf{$\eta=3$}  & \textbf{$\eta=4$}  & \textbf{$\eta=5$} \tabularnewline
\hline 
50  & 12.47  & 12.82  & 13.10  & 13.22  & 13.59 \tabularnewline
80  & 8.12  & 8.37  & 8.55  & 8.68  & 8.77 \tabularnewline
100  & 6.11  & 6.37  & 6.73  & 6.81  & 6.83 \tabularnewline
\hline 
\end{tabular}
\end{table}

\begin{figure}[htbp]
\centering{} \includegraphics[scale=0.42]{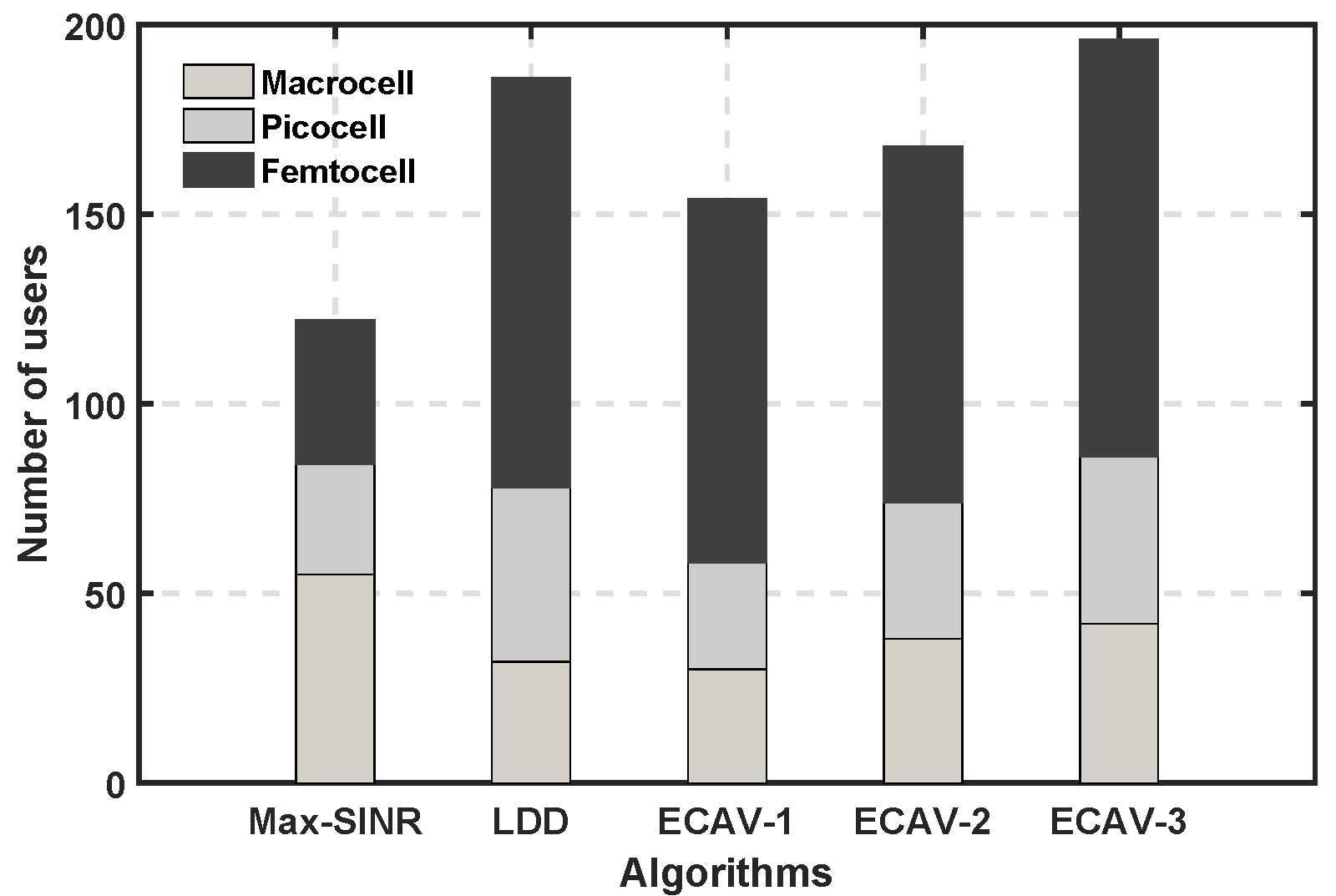}
\caption{The connection between users and BSs according to the allocation scheme
obtained through different algorithms \label{distributed_200} }
\end{figure}

\begin{figure}[htbp]
\centering{} \includegraphics[scale=0.45]{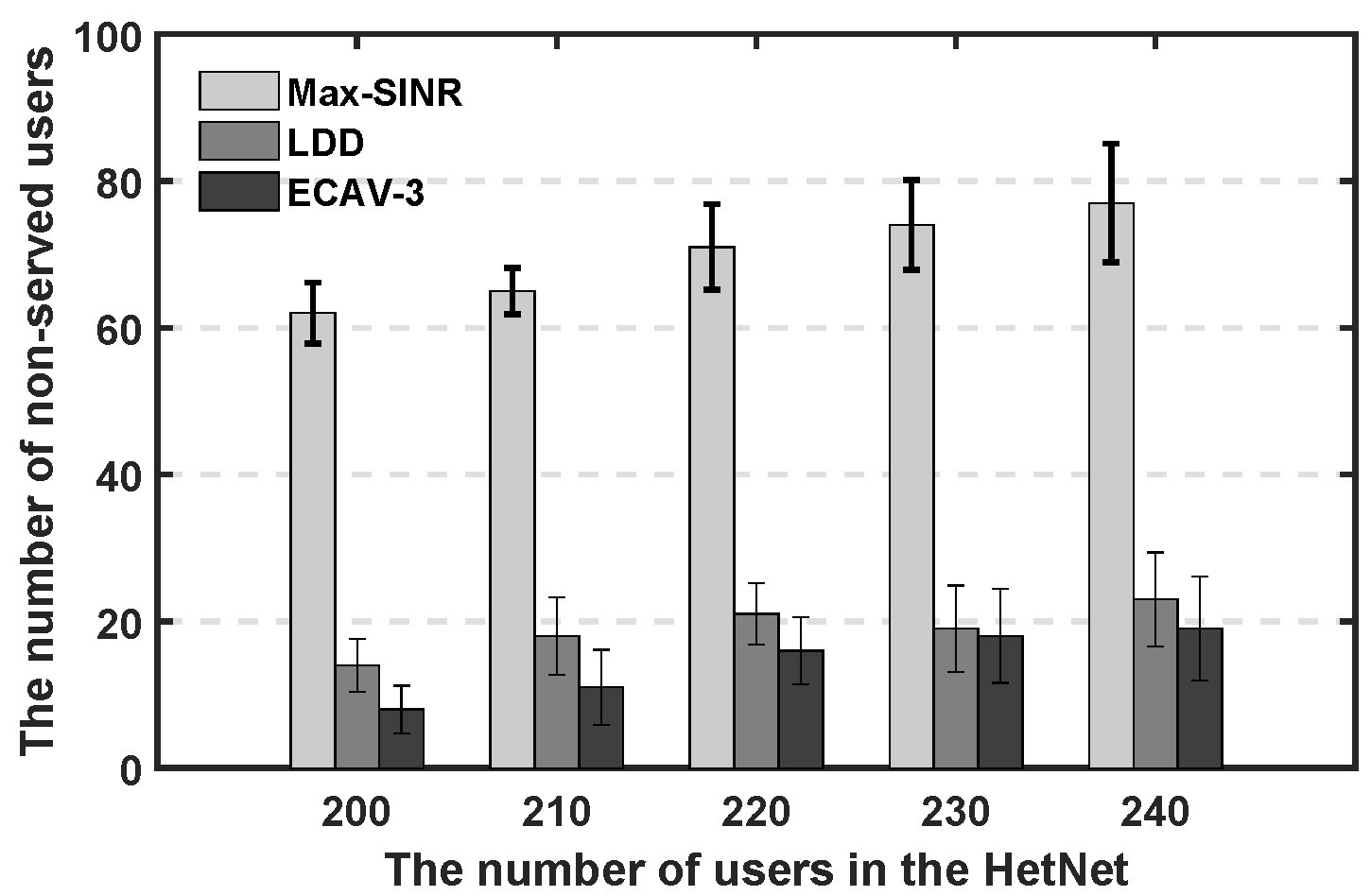} \caption{Non-served users in the HetNet with a different number of users \label{non served} }
\end{figure}

In Fig.\ref{distributed_200}, we check the connection state between
200 users and BSs in different tiers. A phenomenon we can observe
from the figure is that there are more or less some users out of service
even we use different kinds of algorithms. It is not only caused by
the limited resource configured at each BSs, but also related to the
positions of such kinds of users. They are located at the edge of
the square and hardly served by any BS in the system. Further, more
users are served by macro BS in Max-SINR algorithm because a larger
SINR always eixsts between the users and macro BS. As a result, the
total non-served users in Mmax-SINR algorithms are more than the other
two if there is no scheme for allocating the left resource. On the
other hand, the number of non-served users in MC are less than LDD
when $\eta=3$ since the user $\mathcal{U}_{j}$ will select a BS
$\mathcal{B}_{i}$ with the maximal $QI_{ij}$ in each iteration of
the LDD algorithm. In other words, the users prefer to connect with
a BS which can offer better QoS even when more resources are consumed.
Therefore, some BSs have to spend more RBs which leads to the resource
at these BSs being more easily used up.

\begin{figure}[htbp]
\centering{} \includegraphics[scale=0.46]{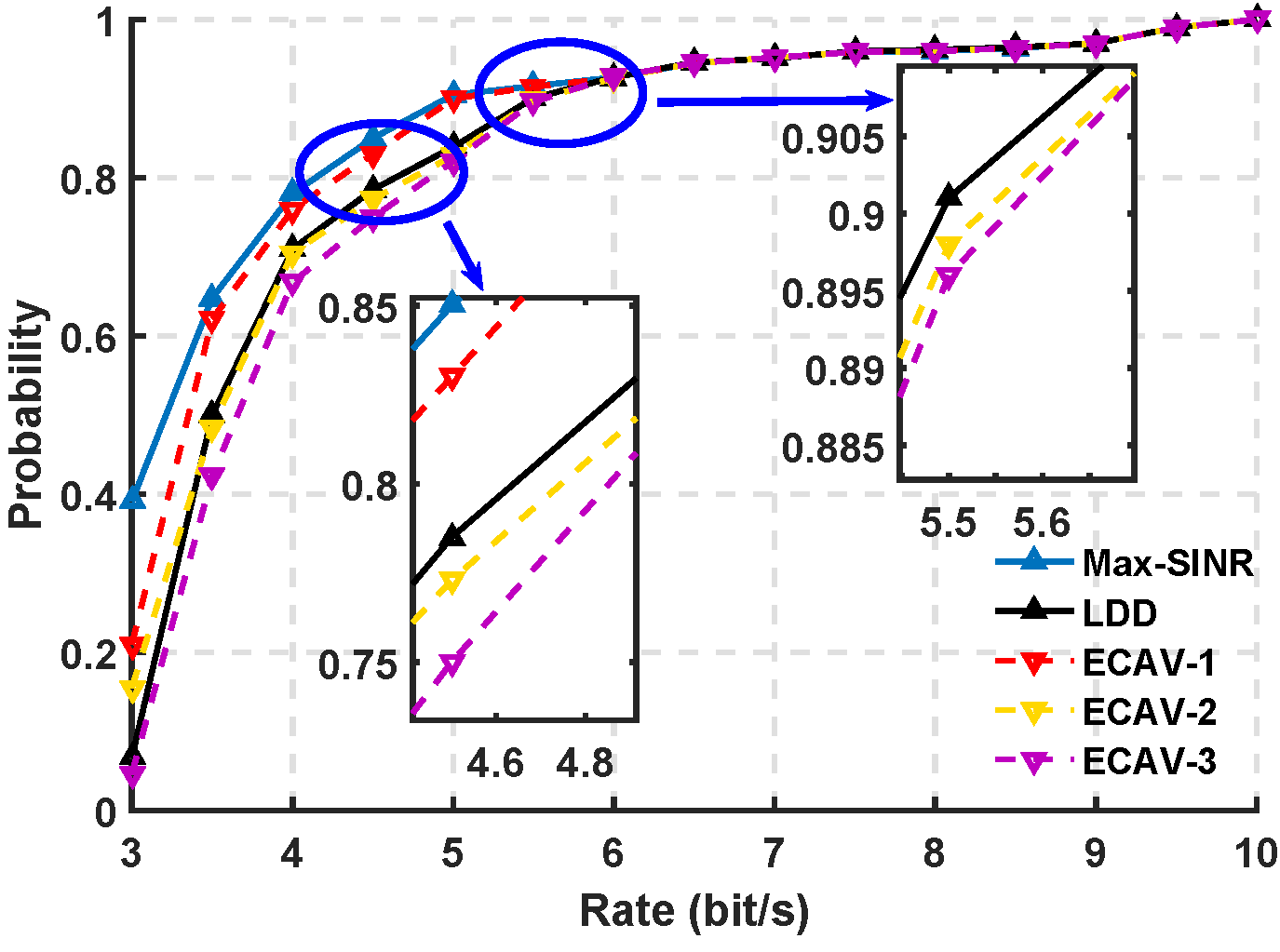} \caption{The CDFs of the rate acheived at users \label{CDF_200}}
\end{figure}

\begin{figure}[htbp]
\centering{} \includegraphics[scale=0.45]{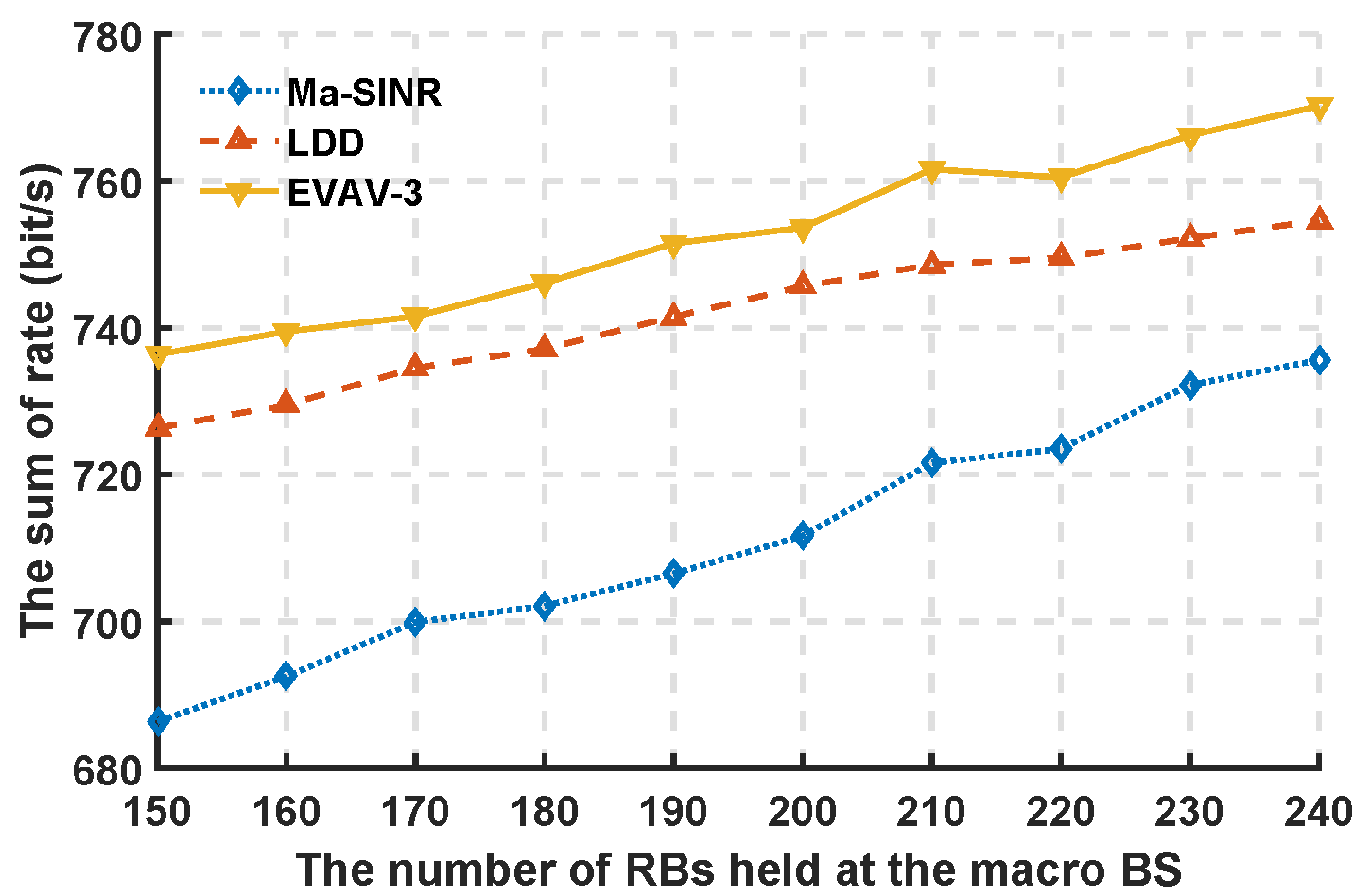} \caption{The total rate against the number of RBs held at the macro base station
\label{number of rb in mac} }
\end{figure}

In Fig.\ref{non served}, we produce a statistic of the number of
non-served users when we change the total number of users configured
in the HetNet. The average number of non-served users for each algorithm
along with the standard deviation is presented in the figure. Compared
with Fig.\ref{distributed_200}, a more clear results imply that more
than 60 (at worst, around 70) non-served users in the Max-SINR algorithm.
The LDD based algorithm comes the second with approximate 20 users.
The best resutls are obtained by the MC algorithm with no more than
20 users even the total users in the HetNet is 240.

In Fig.\ref{CDF_200}, we compare the cumulative distribution function
(CDF) of the rate. The rate of the users seldomly drops below the
threshold (3 bit/s) when we use the distributed algorithms (LDD and
MC based algorithms), while Max-SINR algorithm is unable to satisfy
the rate QoS constraints. Moreover, the rate CDFs of the MC based
algorithm never lie above the corresponding CDFs obtained by implementing
the Max-SINR algorithm (the gap is between $6\%-20\%$). Likewise,
At worst $5\%$ gap eixts between the MC based algorithm and LDD when
we set $\eta=3$.

At last, another intesest observation is made by configurating different
number of RBs at macro BS (Fig.\ref{number of rb in mac}). When we
change the number of RBs from 150 to 250 at macro BS, it is clear
to see that the total rate obtained by LDD is not sensitive to the
variation of the resource hold by macro BS. This result is also related
to the solving process in which two phases are needed when employing
a LDD based algorithm. As we have discuss in the Introduction section,
a solution which can satisfy the basic QoS requirement will be accepted
by the LDD based algorithm. It finally affects the allocation of left
resource at marco BS. As a result, the algorithm easily falls into
the local optima. This problem, to some degree, can be overcome by
the ECAV model since there is only one phase in the model. With the
ECAV model, a constraint satisfied problem is transformed into a constraint
optimizaiton problem. And the advantage of DCOP is successfully applied
into solving user assocation problem.

\section{Conclusion}

\label{conclusion}

An important breakthrough in this paper is that we take the DCOP into
the application of HetNet. More preisely, we propose an ECAV model
along with a parameter $\eta$ to reduce the number of nodes and constraints
in the model. In addition, a markov basesd algorithm is applied to
balance the quality of solution and the time consumed. From experimental
results, we can draw a conclusion that the quality of the solution
obtained by the ECAV-3 model solved with the MC based algorithm is
better than the centralized algorithm, Max-SINR and distributed one
LDD, especially when the number of users increases but they are limited
to the available RBs. In future work, we will extend our research
to the following two aspects:

In some algorithms, like K-opt \cite{2007_kopt} and ADOPT \cite{2005_adopt}
for DCOP, there are already a theoretial analysis on the completeness
of solution. However, it is still a chanllenge job in most research
of DCOP algorithm, like the MC based algorithm proposed in this paper.
Thus, we will explore the quality of the solution assoicated with
different values of $\eta$.

 In practice, the BSs in small cells (like pico/femto BSs) have
properties of plug-and-play. They are generally deployed in a home
or small business where the environment is dynamic. In this way, we
should design a DCOP model which is fit for the variations in the
environment such as the mobility of users and different states (active
or sleep) of BSs. To this end, a stochastic DCOP model can be considered
like the one in .

\end{document}